\documentclass[final,sort&compress,5p,reprint,preprintnumbers]{elsarticle}

\usepackage{graphicx}
\usepackage{dcolumn}
\usepackage{bm}
\usepackage{amsmath}
\usepackage{multirow}
\usepackage{array}
\usepackage{booktabs}

\usepackage[colorlinks=true]{hyperref}

\begin{document}

\title{Development of a Watt-level Gamma-Ray Source based on \\
High-Repetition-Rate Inverse Compton Scattering}

\author[niu]{D. Mihalcea}
\author[radiabeam]{A. Murokh}
\author[niu,apc]{P. Piot\corref{cor1}}
\cortext[cor1]{Corresponding author}
\ead{piot@nicadd.niu.edu}
\author[apc]{J. Ruan}

\address[niu]{Department of Physics and Northern Illinois Center for Accelerator \& Detector Development, \\
Northern Illinois University, DeKalb, IL  60115, USA}
\address[radiabeam]{RadiaBeam Technologies, LLC, Santa Monica, CA 90404, USA}
\address[apc]{Fermi National Accelerator Laboratory, Batavia, IL  60510, USA}
\date{today}

\begin{abstract}
A high-brilliance ($\sim10^{24}$~photon.s$^{-1}$.mm$^{-2}$.mrd$^{-2}$/0.1\%) gamma-ray source  experiment is currently being planned at Fermilab ($E_\gamma\simeq 1.1$~MeV)~[1]. The source implements a high-repetition-rate inverse Compton scattering by colliding electron bunches formed in a $\sim 300$-MeV superconducting linac with a high-intensity laser pulse. This paper describes the design rationale along with some of technical challenges associated to producing high-repetition-rate collision. The expected performances of the gamma-ray source are also presented.
\end{abstract}
%
\maketitle

\section{Introduction}
High-flux, quasi-monochromatic, $\gamma$-ray sources could have widespread range of applications including in Nuclear Astrophysics, Elementary Particle Physics and Homeland security.  In the later class of application, developing compact gamma-ray source capable of producing large flux could enable the rapid screening of cargos for fissile material detection. The need for monochromatic $\gamma$ rays along with the requirement for a small-footprint source have motivated the exploration of particle-accelerator-based sources employing inverse Compton scattering (ICS)~\cite{Aru,Kul}. This development path is further supported by the increasing number of compact GeV-class electron sources based on laser-plasma wakefield accelerators (LPAs) available at various laboratories worldwide~\cite{lpa}. LPAs have so far been employed to generate $\gamma$ rays with impressive brilliance but with restricted photon flux due to their low operating frequencies (typically 10~Hz) limited by the current solid-state-laser technologies~\cite{loa,nebraska}. Such a limitation of the LPAs is currently being addressed by several groups while the development of high-repetition rate interaction region could be performed at available state-of-the-art linear accelerators.  Based on this observation, a collaboration between Fermilab, Northern Illinois University and RadiaBeam LLC is currently designing an experiment aimed at demonstrating high-repetition-rate ICS using the Fermilab Accelerator Science \& Technology (FAST) facility based on a superconducting linear accelerator~\cite{Leib}. The present paper summarizes the design rationale and expected performances of the source under design. 
\begin{figure}[bb!!]
\begin{center}
\includegraphics[width=0.95\linewidth]{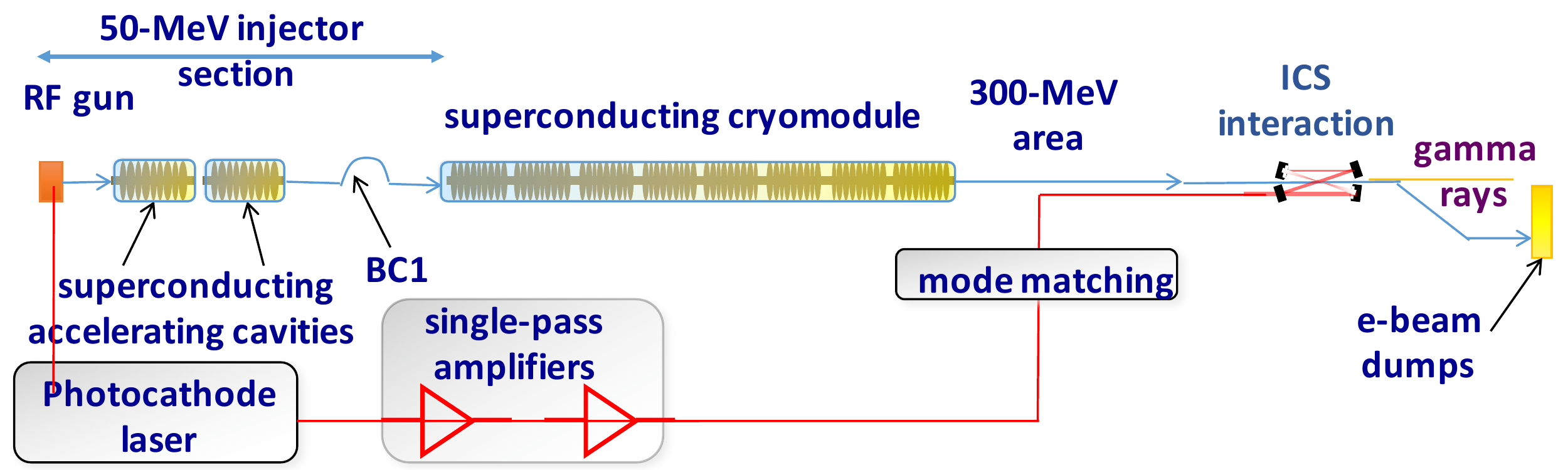}
\end{center}
\vspace{-1.0em}
\caption{\label{fig:overview} Overview of the $\gamma$-ray source under consideration at the FAST facility. The "BC1" and "ICS" labels respectively refer to the bunch compressor and inverse-Compton scattering. }
\end{figure}

\section{Overview of the $\gamma$-ray source}
Figure~\ref{fig:overview} provides an overview of the source concept to be driven by the 300-MeV electron beam available at the FAST facility. In brief the electron bunches are produced via photoemission from a semiconductor photocathode located in a $1+\frac{1}{2}$-cell radiofrequency (RF) gun. The bunches are photoemitted by impinging an ultraviolet (UV) laser pulse obtained via frequency quadrupling of an amplified infrared (IR) laser pulse ($\lambda=1053$~nm) produced from a Nd:YLF laser system. The formed bunches are further accelerated by two TESLA-type superconducting RF (SRF) cavities. The RF gun and cavities operate at 1.3~GHz and are capable of forming 1-ms bunch trains at 5-Hz frequency and containing up to 3000~bunches (corresponding to a 3-MHz bunch frequency). The 50~MeV beam can be manipulated and diagnosed before injection in an international-linear-collider- (ILC) type accelerating cryomodule. The cryomodule incorporates 8~SRF cavities with demonstrated average accelerating gradient of $\bar{G}\simeq 31.5$~MeV/m resulting in a maximum beam energy of $\sim 300$~MeV. The beam is further transported along a $\sim 70$~m beamline and finally directed to the ICS interaction point (IP). Downstream of the IP, the electron beam is then sent to a high-power beam dump.

The laser pulse used in the ICS interaction is derived from the photocathode laser system. The unspent IR laser energy downstream of the IR-to-UV conversion process is conditioned, further amplified, and directed to a passive coherent enhancement cavity for final amplification. The laser pulses collides head-on with the electron bunches with a collision frequency of 3~MHz within the 1-ms train thereby yielding $\gamma$-ray pulses with similar format. The energy of the backscattered $\gamma$-ray photon is $E_{s}\simeq \hbar \omega_s$ where $ \hbar$ is the reduced Planck's constant, and the upshifted backscattered-photon frequency  is
$
\omega_s(\theta) =\frac{4\gamma^2 \omega_l}{1+a_0^2/2+\gamma^2\theta^2}, 
$
with $\hbar\omega_l \simeq1.2$~eV and $a_0$ being respectively the laser photon energy and normalized potential, and $\gamma$ the electron-beam Lorentz factor. The angle $\theta$ is the direction of observation referenced to the electron-beam direction.  Therefore the 300-MeV beam available at FAST will support the generation of $\gamma$ rays with $E_s\le 1.5$~MeV. Considering a laser pulse energy  ${\cal E}_l \sim0.5$~J and a focused transverse size $\sigma_l\simeq 30$~$\mu$m,  we obtain $a_0^2/2 \simeq 7.6\times 10^{-3} \ll 1$ thereby confirming that nonlinear effects are insignificant. Owing to its narrow bandwidth ($\delta\lambda\sim 0.2$~nm) the laser transform-limited rms pulse duration is 3~ps close to the measured value of 3.8~ps.

\section{Expected Performances}
Applications of $\gamma$ rays generally require high photon flux, narrow spectral bandwidth and high brilliance. The brilliance can be expressed as a function of the laser and electron- beams parameters \cite{Brown2} as 
\begin{eqnarray}
{\cal B}_s \propto \frac{N_l}{\sigma_l^2} \gamma^2 \frac{N_e}{\tau_e \epsilon_{n,x}^2},
\label{eqn:Bs}
\end{eqnarray}
where $N_l$ (resp. $N_e$) are the number of photons (resp. electron) in the laser pulse (resp. electron bunch), $\tau_e$ the electron-bunch duration and $\epsilon_{n,x}$ its normalized transverse emittance. Likewise, the backscattered-photon dose $N_s$ and relative spectral bandwidth of the scattered pulse $\delta\omega_s/\omega_s$ can be respectively parameterized as 
\begin{eqnarray}
N_s \approx \frac{N_e N_l \sigma_T}{2 \pi (\sigma_e^2 + \sigma_l^2)}, \mbox{and~} \frac{\delta \omega_s}{\omega_s} \approx \frac{\epsilon_{n,x}^2}{\sigma_e^2}
\label{eqn:dose},
\end{eqnarray}
where $\sigma_e$ is the electron-beam transverse rms size and $\sigma_T$ the Thompson's cross section. Other sources contributing to  $\delta\omega_s/\omega_s$ includes the laser spectral bandwidth and fraction momentum spread of the electron beam the cumulative contribution of these two effect is found to be negligible ($\sim 0.2$~\%) for our operating regime. The electron-beam emittance is the dominant contribution to the scattered-radiation bandwidth. It should finally be pointed out that Eqs~\ref{eqn:Bs} and~\ref{eqn:dose} assume the laser and electron beams are cylindrically symmetric. \\

We modeled the ICS process using the program {\sc compton}~\cite{Brown} which provides a 3D treatment of the laser-electron instruction in the time and frequency domains. The software also includes the nonlinear corrections to the Thompson cross section to correctly account for the electron recoil at the time scale of the interaction process. To gain further confidence in our modeling, we also  employed the program {\sc cain}~\cite{CAIN} which is based on the more general Klein-Nishina cross-section for Compton-scattering process and consequently accounts for the quantum-electrodynamical (QED) effects. In our case, taking $\gamma \approx 500$, we find that the laser photon energy in electron-beam frame verifies $2 \hbar  \gamma \omega_l \approx 0.6$~keV$\ll m_ec^2 = 511$~keV (where $m_ec^2$ is the electron's rest mass) so that QED effects are expected to be insignificant. The codes  {\sc compton} and {\sc cain} were benchmarked and found to be in reasonable agreement over the considered range of parameters~\cite{mihalceaAAC16}. \\

\begin{figure}[hhh!!!!!!!]
\begin{center}
\includegraphics[width=0.99\linewidth]{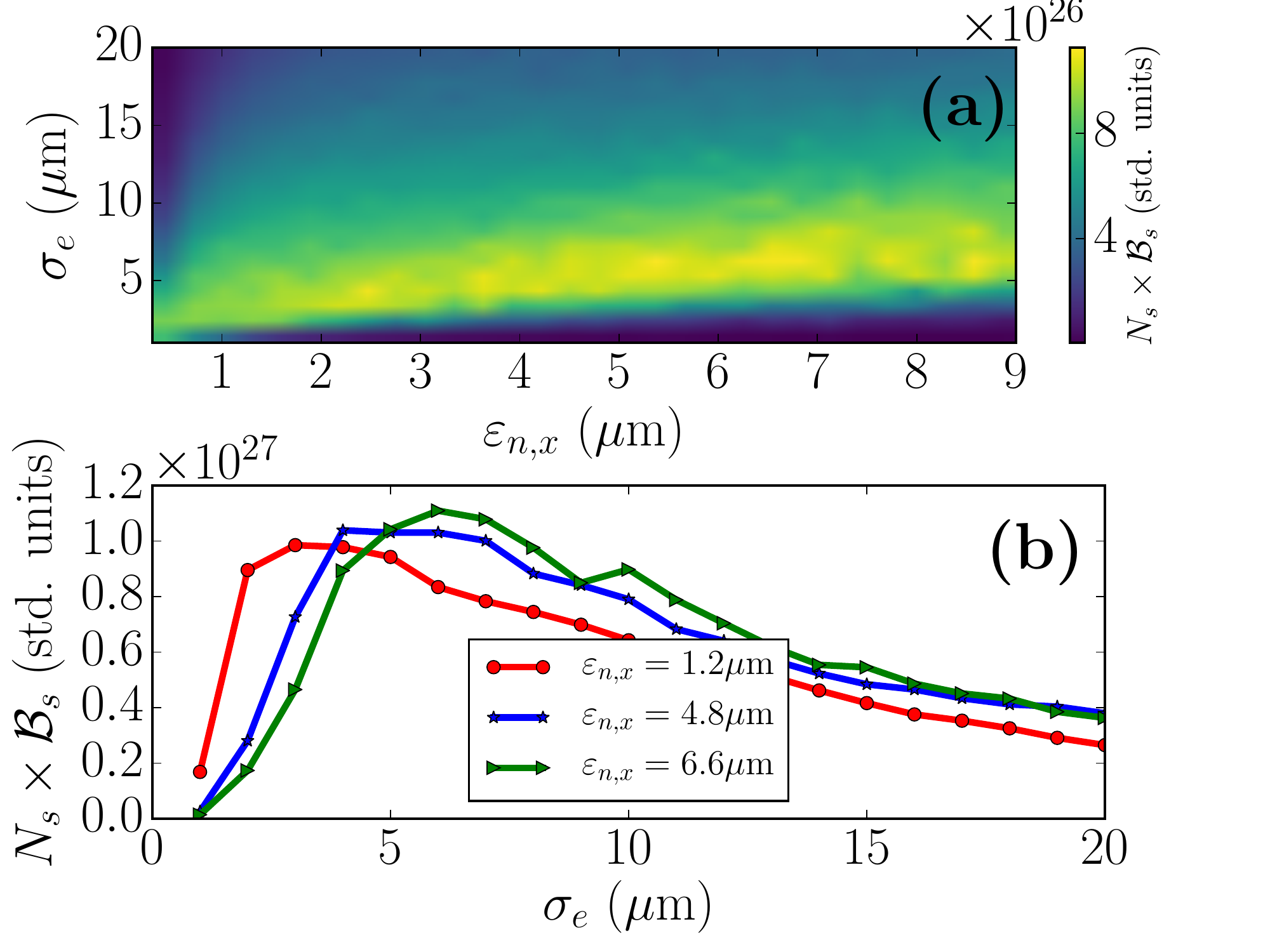}
\end{center}
\vspace{-1.0em}
\caption{Scatter plot of the product $N_s \times B_s$ as a function of electron beam transverse emittance and transverse size (a) and evolution of $N_s \times B_s$ as a function of electron beam transverse size for three values of the emittance (b).}
\label{AX}
\end{figure}

The {\sc compton} program was used to devise electron-beam and laser parameters such that scattered-pulse brightness, dose and bandwidth are optimized using Eq.~\ref{eqn:Bs} and~\ref{eqn:dose} as guidelines. Additionally, previous beam dynamics simulation~\cite{PP} indicated that the transverse beam emittance and charge are related via the scaling relation $Q \mbox{[nC]} \approx
0.34 \varepsilon_{nx}^{1.45} \mbox{[$\mu$m]}$. In our optimization we took a more conservative relationship $Q \mbox{[nC]} \approx 3  \varepsilon_{n,x}\mbox{[$\mu$m]} $ which approximates well the the previous equation for $Q \in [0.1,3]$~nC. This empirical relationship reduces the quantity $A_s \equiv N_s \times B_s \propto \frac{1}{\sigma_e^2 +\sigma_l^2}$  to be solely dependent on the transverse sizes.

The quantity $A_s$ was computed as a function of beam emittance and electron-beam size; see Fig.~\ref{AX}. For these calculations, the brightness and dose are computed by only considering scattered photons with relative frequency $\frac{\delta \omega_s}{\omega_s}\le 1$\%. Figure~\ref{AX} suggests a range of beam parameter that maximizes $A_s$. It also indicate that small transverse  beam sizes ($\sigma_e<3 \mu$m) are not favored as  $A_s$ decreases. In this range of value the scaling relations Eqs.~\ref{eqn:Bs} and~\ref{eqn:dose} do not hold anymore as diffractive effects become omportant and the spectral bandwidth of the radiation significantly increases resulting in a drastic reduction of number of photon withins $\frac{\delta \omega_s}{\omega_s}\le 1$\%. A possible working point devised from Fig.~\ref{AX} sets $\sigma_e \simeq 5\mu$m which maximizes $A_s$ over a large range of electron-beam emittance. Additionally, forcing the bandwidth to remain at $\frac{\delta \omega_s}{\omega_s}\le 2$~\% sets the emittance to  $\varepsilon_{n,x}\le 0.6$~$\mu$m. According to the empirical charge-emittance relationship mentioned above, such an emittance corresponds to an electron-bunch charge $Q\simeq 200$~pC.
\begin{table}[htb]
\centering
\begin{tabular}{l l l l}
\hline
window  & 1~\%    & 3~\% & 100~\%  \\
\hline
${\cal B}_s$ & $2.7 \times10^{20}$ & $2.1 \times 10^{20}$ & $1.1 \times 10^{20}$ \\
$N_s$  (phot.) & $5.4 \times 10^5$ & $2.0 \times 10^6$ & $4.8 \times 10^7$ \\
$\delta\omega_s/\omega_s$ (\%) & 0.25 & 0.82 & 2.2 \\
\hline
\end{tabular}
\caption{Optimized single-bunch $\gamma$-ray parameters for three different energy windows where 1, 3 and 100\% of the scattering photon are considered. The listed parameters are simulated for a single-bunch interaction, ${\cal B}_s$ and $N_s$ should be multiplied by 15,000 in order to scaled the parameters to the full bunch train. }
\label{param}
\end{table}

\begin{figure}[hhhhh!!]
\begin{center}
\includegraphics[width=.99\linewidth]{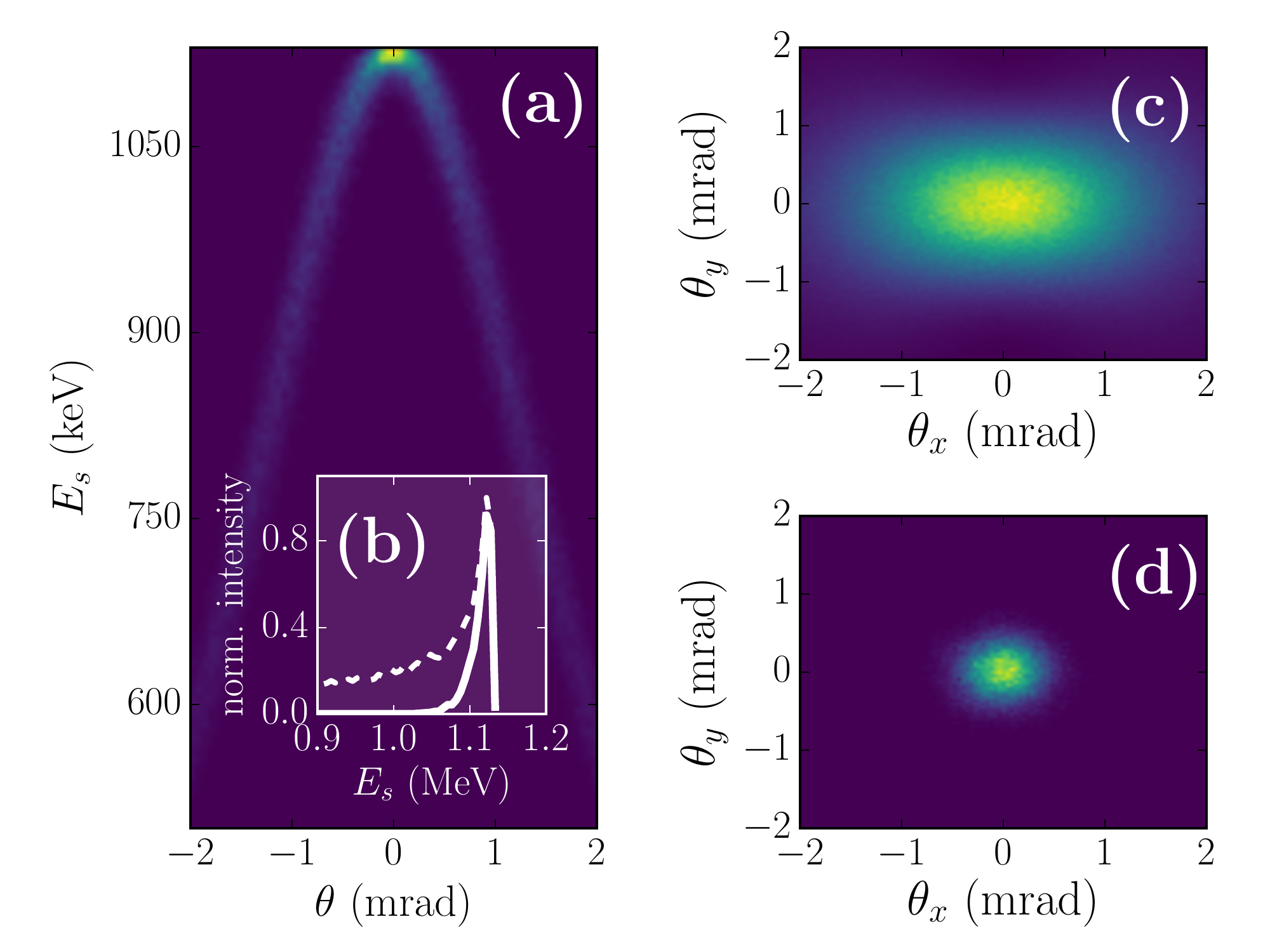}
\end{center}
\vspace{-1.0em}
\caption{\label{fig:final} Simulated density plots associated to the angular-spectral (a) and transverse (b,c) distributions of  $\gamma$-ray generated using the set of parameters in Tab.~\ref{tab:elaser}. Plot (c) is computed when all the generated photons are taken into account while plot (d) only consider the 1\% most-energetic photons. Inset (b) shows the corresponding integrated spectra as dashed (all photons) and solid (1\% most-energetic photons) lines.}
\end{figure}

Table~\ref{param} gathers the key parameters of the scattered photon pulse evaluated for three given percentiles of scattered photons (1~\%, 3~\% and 100~\%).  Such parameters are relevant in practice as only a small fraction of the scattered photons (i.e. the most energetic) are employed in front-end experiments. The selection is typically accomplished using transverse collimators exploiting the correlation between photon energy and angular spread (red shifting occurs off axis). Table~\ref{param} confirms that the peak brightness and bandwidth improve while the dose worsens as the percentile of considered photon decreases.  It should be pointed out that the Tab.~\ref{param} provides the scattered pulse properties for a single-bunch collision. The brightness and dose should be multiplied by 15,000 in order to scale these parameters to the multi-bunch collision case (i.e. when 5-Hz bunch train composed of 3000 bunches are employed). Owing to the excellent multi-bunch stability of superconducting linac the spectral bandwidth remains unaffected in the multi-bunch operation case. 
\begin{table}[htb]
\centering
\begin{tabular}{lclc}
\toprule
  \multicolumn{2}{c}{electron beam} & \multicolumn{2}{c}{laser pulse}\\
\midrule
 beam energy (MeV)     & 250.0      & wavelength (nm)  & 1053 \\
 beam charge (pC)        & 200.0      & pulse energy (J)   & 0.5 \\
 energy spread (\%)      & 0.1        & bandwidth   (nm)  & 0.2 \\
 duration (ps)                & 3.0          & duration   (ps)  & 3.0 \\
 beam size ($\mu$m)   & 5.0          & spot size ($\mu$m)   & 15.0\\
 emittances ($\mu$m)  &  0.6         &   &  \\
\bottomrule
\end{tabular}
\caption{Summary of electron-beam and laser-pulse parameters used to produce Tab.~\ref{param} and Fig.~\ref{fig:final}. The duration, spot size, energy spread and bandwidth are given as rms values. } 
\label{tab:elaser}
\end{table}

Table~\ref{tab:elaser} provides the final set of electron- and laser-beam parameters while Fig.~\ref{fig:final} summarizes the angular-spectral and transverse distributions and also provides spectra computed for the full photon ensemble and the 1\% most energetic photon set.

\section{Interaction region}
\begin{figure}[hhhh!!]
\begin{center}
\includegraphics[width=0.96\linewidth]{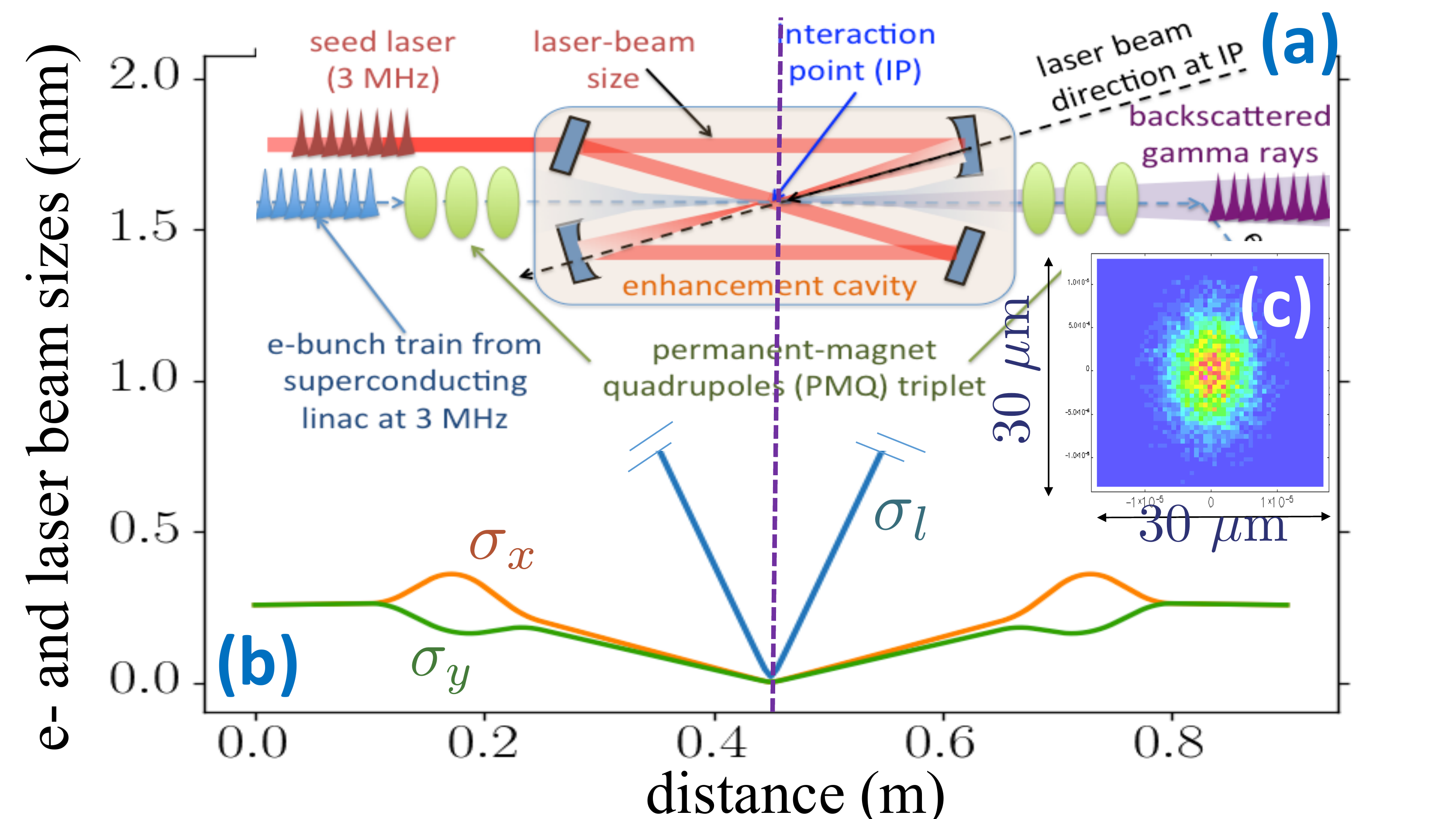}
\end{center}
\vspace{-1.0em}
\caption{\label{fig:ffoc} Overview of the final-focus area (a) with associated rms beam sizes evolution (b). In (b) $\sigma_{x,y}$ are the electron-beam transverse beam sizes while $\sigma_l$ denotes the laser spot size evolution. The origin of the distance ordinate is $z=120$~m from the photocathode surface. Inset (c) shows a density plot of the electron beam transverse density at the IP [located at distance of 0.43~m in (b)].}
\end{figure}

At the IP, final-focus high-gradient quadrupole magnets are needed to produce tight electron-beam waist at the interaction point (IP), the optical enhancement cavity and some diagnostics. The magnetic lattice combines two quadrupole-magnetic triplets (each composed of three magnets) respectively located upstream and downstream of the IP; see Fig.~\ref{fig:ffoc}(a). The upstream triplet is used to focus the beam and match its transverse size to the laser size while the downstream triplet rematches the electron beam parameters to the downstream accelerator beamline for transport up to the beam dump. The focusing magnets are Halbach-type permanent-magnet quadrupoles (PMQs)~\cite{pmq} capable of producing the required field gradient of $g\simeq300$~T.m$^{-1}$. The use of such strong PMQ yields a meter-scale interaction region compatible with possible future coupling to LPAs. An example of evolution of the electron-beam transverse size appears in Fig.~\ref{fig:ffoc}(b) and the associated transverse distribution at the IP is shown Fig.~\ref{fig:ffoc}(c). The PMQs are transversely aligned using an in-situ laser-based alignment, whereas their axial positions is adjustable to provide some flexibility. Further control on the interaction-point waist is provided by electromagnetic quadrupole magnets available upstream (and downstream) of the interaction chamber.

As previously pointed out, the laser beam used for the ICS process originates from the IR photocathode laser system. Our current approach is to condition and to amplify the IR pulses using single-pass amplifiers to bring the laser pulse energy from $\sim 50$~$\mu$J to $\sim 5$~mJ eventually and then to further increase the pulse energy to Joule level using a passive enhancement cavity. This type of cavities were intensively studied in the recent years and intensity amplification factors of several thousands were reported~\cite{Shim,Bri}. An amplified IR laser pulse enter in such a cavity, illustrated in Fig.~\ref{fig:ffoc}(a), from the left side and undergo an integer number of round trips, a new laser pulse arrives at the cavity entrance. Assuming a optimal mode matching the amplitude of the two pulses add up coherently. Ideally, the intensity build-up is only limited by the loses in the mirrors. The process is repeated with the subsequent IR pulses. Once the steady state is reached, the intra-cavity pulse intensity is $I = \frac{1 -R_1}{1 - \sqrt{(R_1 R_2)^n} } I_L$ where $R_i$ ($i=1,2$) are the reflectivity coefficients of the two cavity mirrors, $n$ is the number of roundtrips performed by the laser pulse inside the cavity during the time interval between consecutive bunches coming from the laser system and $I_L$ is the intensity of the laser pulses just before the entrance in the cavity.  Consider the case of a conventional Perot-Fabry resonator with commercially-available mirrors with $R_1 = 0.999$, $R_2 = 0.99995$ and $n = 1$ the total gain is $\Gamma\equiv \frac{I}{I_L} \approx 3,626$. In such a two-mirror configuration, the injection mirror, must have a much lower reflectivity in order to allow the outside laser pulses to couple to the cavity. This later feature results in significant losses when a large $n$ is required. In our case where the pulse repetition rate is 3~MHz (corresponding to a pulse separation of $\sim 100$~m), requiring the cavity to fit within a 2-m footprint would imply $n=25$ yielding a poor gain  $\Gamma \approx 6$. Consequently an alternative approach is required. One candidate being explored is a Herriott cell~\cite{herriott} combined with a four-mirror bow-tie cavity. Calculations performed for such a configuration indicate a gain $\Gamma \sim100$ is attainable but further optimization is needed.  

\section{Summary}
The coupling of an electron beam originating from a 250-MeV SRF pulsed linac with a high-repetition rate infrared laser is expected to provide $\gamma$ rays with maximum energy up to $E_s\sim 1.1$~MeV, average brilliance of $\langle{\cal B}_s\rangle \simeq 4.4 \times10^{24}$~photon.s$^{-1}$.mm$^{-2}$.mrd$^{-2}$/0.1\%, photon flux of $\langle N_s\rangle \simeq 8 \times 10^9$~photon/sec and relative spectral bandwidth of $\delta\omega_s/\omega_s \simeq 2.5\times10^{-3}$.  These predicted parameters corresponds to an average power of $\sim6$~mW and are expected to foster a wide range of applications such as fissile-material detection, exploration and development of gamma-ray optics, or possibly enable the quantum control of nuclear states. 

We are grateful to Dr. W. Brown for sharing his inverse Compton-scattering simulation software {\sc compton}. This work was sponsored by the DNDO award 2015-DN-077-ARI094 to Northern Illinois University. Fermilab is operated by Fermi Research Alliance, LLC. for the U.S. Department of Energy under contract DE-AC02-07CH11359.

\end{document}